\magnification=1200
\overfullrule=0pt
\baselineskip=20pt
\parskip=0pt
\def\dag{\dagger}
\def\del{\partial}

\def\a{\alpha}     
\def\b{\beta}      
\def\g{\gamma}     
\def\d{\delta}     
\def\e{\epsilon}   
      
\def\j{\eta}       
\def\q{\theta}

\def\n{\nu}        
\def\x{\xi}        
          
\def\p{\pi}        \def\P{\Pi}
\def\r{\rho}       
\def\s{\sigma}

\def\y{\psi}       \def\Y{\mit\Psi}
\def\w{\omega}     \def\W{\mit\Omega}
\def\br{\langle}
\def\ke{\rangle}
\def\ve{\vert}
\def\ydag{\y^{\dag }}
\def\ytdag{\tilde\y^\dag }
\def\yt{\tilde\y }
\def\rhoh{\hat{\rho }}

\def\zbar{\bar{z}}
{\settabs 5 \columns
\+&&&&CCNY/HEP/01/04\cr}
\bigskip
\centerline{\bf Bulk and edge excitations of a $\nu =1$ Hall ferromagnet}
\bigskip\bigskip
\centerline{Rashmi Ray$^1$ and B. Sakita$^2$}
\bigskip
\centerline{ American Physical Society, One Research Road, Ridge NY 11981$^1$}
\bigskip
\centerline{ Physics Department, City College of the City University of New
York}
\centerline{ New York, NY 10031$^2$}
\bigskip
\centerline{\bf Abstract}
\bigskip
In this article, we shall focus on the collective
dynamics of the fermions in a $\nu = 1$ quantum Hall droplet.
Specifically, we propose to look at the quantum Hall ferromagnet.
In this system, the electron spins are ordered in the ground state
due to the exchange part of the Coulomb interaction and the Pauli exclusion
principle. The low energy excitations are ferromagnetic magnons.
In order to obtain an effective Lagrangian for these magnons,
we shall introduce bosonic collective coordinates
in the Hilbert space of many-fermion systems. 
These collective coordinates describe a part of the fermionic Hilbert space.
Using this technique, we shall interpret the magnons as bosonic collective 
excitations in the Hilbert space of the many-electron Hall system.
Furthermore, by considering a Hall droplet of finite extent, we shall also obtain
the effective Lagrangian governing the spin collective excitations at the
edge of the sample.
\vfill
\noindent{$^1$E-mail address: ray@aps.org.}

\noindent{$^2$E-mail address: sakita@sci.ccny.cuny.edu.}
\vfill\eject
\centerline{\bf I. Introduction}
\bigskip
Quantum Hall ferromagnets have generated a considerable amount of interest
in recent years.
The fact that Hall states should exhibit ferromagnetism is somewhat 
surprising. In conventional Hall systems, the electron spins are
aligned by the strong external magnetic field. Thus, in the absence of
spontaneous alignment, these Hall states do not qualify as 
ferromagnetic states. However, particular samples, where the coupling
of the electronic spins to the external magnetic field is
negligible for a variety of reasons, do exhibit a spontaneous alignment
of the spins. These are the Hall ferromagnets [1],[2],[3].

Traditionally, the integer and the fractional 
Hall effects are distinguished by the origin of the gap in the single 
particle spectrum.  
In the integer effects, the gap
which is equal to the cyclotron gap, is produced
by the energy difference between the successive Landau levels.
On the other hand, in the fractional effects, the gap is produced
due to inter-electronic interactions in the strongly correlated Hall fluid.

There is, however, a scenario, where despite integer filling, the gap in 
the single particle spectrum is due to inter-electron interactions. 
For instance, in GaAs, the effective mass in the conduction band, which
appears in the expression for the cyclotron gap, is much smaller than
the actual mass of the electron, which appears in the expression for
the Zeeman gap, through the Bohr magneton $\mu_{B}$. Thus the cyclotron
gap increases effectively by a factor of $\sim 14$. Spin-orbit 
scattering reduces the effective $g$ factor by a factor of $\sim 5$.
Thus the Zeeman energy is vanishingly small compared to the cyclotron
energy (a factor of $\sim 70$ smaller)[1]. 

Imagine that, starting from the value of 2, the $g$ factor in the sample can be
gradually brought down to zero. Initially, for $g=2$, the $\nu =1$ state
described a state with a uniform density of ${{B}\over{2\p }}$ electrons
per unit area, with each electron  in the spin ``up" state, say.
With the reduction of the Zeeman gap, one would expect this state to be 
rendered unstable as the two spin orientations gradually become degenerate.
Experimentally, however, a gap is observed, indicating the $\nu =1$
state remains stable [4].

A rather heuristic argument may be adduced to explain this phenomenon.
In the lowest Landau level, the kinetic energy of the electrons has a fixed
value. Thus, to minimise the Coulomb energy, the spatial part of the
many-electron wave function should be totally antisymmetric. This in turn
will require, by the Pauli principle, that the spin part of the wave function
be completely symmetric. Thus, the Coulomb interaction causes the
spins to align and is instrumental in producing the observed gap which
stabilises the $\n =1$ state in the absence of the Zeeman coupling. 
The state is genuinely ferromagnetic as the spins are spontaneously
aligned. In this situation, the conventional distinction between the integer and
the fractional effects becomes somewhat blurred.

Since the ground state exhibits the spontaneous breaking of the
original global spin $SU(2)$ symmetry (for $g=0$) down to a $u(1)$
symmetry, the excitation spectrum must contain gapless Goldstone
bosons, namely, the ferromagnetic magnons. The effective Lagrangian
governing the dynamics of the magnons has been obtained previously
in a variety of ways [1],[2],[5].

It is well known that the excitations located at the edges of finite
Hall samples play a crucial role in the physics of these samples.
The so-called edge states have been studied in considerable depth for
the integer and the fractional effects [6]. 
In contrast, despite some seminal
work [7] on the edge excitatitons of ferromagnet Hall samples,
a systematic derivation of the effective Lagrangian 
governing these excitations seems to be in order.

In this article, we propose to use a recently developed technique [8]
for introducing bosonic collective variables as coordinates in 
many-fermion Hilbert spaces, in studying the bulk and the edge excitations
of the $\nu =1$ Hall ferromagnet.
\bigskip
\centerline{\bf I. Bosonic Effective Lagrangian for Fermions}
\bigskip
In this section, we shall review a method, originally developed in [8], of introducing bosonic
collective coordinates in the Hilbert space of many-fermion systems.
We shall further obtain, starting from
the second-quantised fermionic Lagrangian, an effective Lagrangian
governing the bosonic collective coordinates.

Let us consider a system of $N$ fermions, free to reside at 
$K$ sites, with $K \geq N$. Let us denote the sites by $\alpha $ 
($\alpha = 1,2, \cdots ,K$). 
An $N$-body fermionic Fock state is given by 
$$
\vert \alpha_{1}, \alpha_{2}, \cdots , \alpha_{N} \ke = 
\ydag_{\a_{1}} \ydag_{\a_{2}}\cdots 
\ydag_{\a_{N}}\vert 0 \ke . \eqno (1.1)
$$
where $\ydag_{\a }$ are the fermionic creation operators 
satisfying the standard anti-commutation relation 
$\{ \psi_{\alpha },\ydag_{\b } \} = \delta_{\alpha \beta } $
and where $\vert 0 \ke $ is the fermionic vacuum state.
The number of independent Fock states
describing this system is given by 
$$
D_{f} = {{K !}\over{{(K-N)!}{N!}}}\eqno (1.2).
$$
Let $A$ denote a set of $N$ indices $\a_{1},\a_{2}, \cdots \a_{N}$.
A general state vector $\vert \Y \ke $ belonging to the
$N$-particle Fock space is given by
$$
\eqalignno{
\vert \Y \ke &= \ \sum_{\a_{1}, \a_{2}, \cdots \a_{N}}
{{1}\over{\sqrt{N!}}}\ \vert \a_{1}, \a_{2}, \cdots \a_{N} \ke 
\Y_{\a_{1}, \a_{2}, \cdots \a_{N}} \cr 
&\equiv \sum_{A} \vert A \ke \Y_{A} & (1.3) \cr 
}
$$
where $\Y_{\a_{1}, \a_{2}, \cdots \a_{N}}$ is a totally anti-symmetric 
complex field with $N$ indices. We normalise it to 
$\br \Y \vert \Y \ke =$$1 $.
Thus,
$$
\sum_{\a_{1}, \a_{2}, \cdots \a_{N}} {{1}\over{N!}}
\vert \Y_{\a_{1}, \a_{2}, \cdots \a_{N}}\vert^{2} = \sum_{A} \vert \Y_{A}
\vert^{2} =  1 . \eqno (1.4)
$$
Based on $(1.3)$, we may look upon $\Y_{A}$ as the coordinates of
the fermionic Hilbert space.

>From simple group theoretic considerations, we also know that 
the completely antisymmetric irreducible
representation of the group $SU(K)$, in terms of antisymmetric
tensors with $N$ indices, has a dimensionality equal to $D_{f}$.
Thus, the group $SU(K)$ acts naturally on the $
\Y_{\a_{1}, \a_{2}, \cdots \a_{N}}$. Namely,
$$
\Psi_{\a_1 \a_2 \cdots \a_N} = {{1}\over{{\sqrt{N!}}}}
\Phi_{\b_1 \b_2 \cdots \b_N}u_{\a_1 \b_1}u_{\a_2 \b_2}\cdots u_{\a_N 
\b_N}\eqno (1.5)
$$
where the indices $\a , \b $ run from 1 to $K$.
The transformation $(1.5)$ is equivalently written as
$$
\Y_{A} = \sum_{B} {\cal D}_{AB}(u) \Phi_{B} , \eqno (1.6)
$$
where ${\cal D}(u)$ is the totally anti-symmetric irreducible 
representation of $SU(K)$, of dimensionality $D_{f}$.

Using $(1.5)$, we perform a change of coordinates from $\Y_{A}$
to $u,\Phi_{B}$. The $u$ are the parameters of $SU(K)$ and we call
them {\it collective coordinates}. If we restrict the range of $B$ to
less than $D_{f}-{{K^{2}}\over{2}}$, the new coordinates do not
cover the entire Hilbert space. Nonetheless, if we choose the 
restricted $\Phi_{B}$ judiciously, we could hope to include the
important collective states in the truncated Hilbert space.
In this paper, we restrict $B$ to one single value, namely,
a set of $1,2,\cdots N$. We set
$$
\Phi_{\b_{1} \b_{2} \cdots \b_{N}} = \e_{\b_{1} \b_{2} \cdots \b_{N}}, \eqno (1.7)
$$
where $\e_{\b_{1} \b_{2} \cdots \b_{N}}$ is the Levi-Civita symbol
for the set $1,2,\cdots N$. 
Substituting $(1.7)$ into $(1.5)$ and thence into $(1.3)$, we obtain
$\vert \Y \ke \equiv \vert u \ke $ where
$$
\vert u \ke  = \prod^{N}_{\a =1}\ytdag_{\a } \vert 0 \ke  \eqno (1.8)
$$
with
$$
\ytdag_{\a } = (\ydag u)_{\a },\ \{ \yt_{\a },\ytdag_{\b }\} =\d_{\a \b }
. \eqno (1.9)
$$
We get $\br u \vert u \ke =1$. The state $\ve u \ke $ is obviously
constructed by filling up $N$ sites sequentially with fermions created
by $\ytdag_{\a }$.

Let us define an operator ${\cal P}$ as
$$
{\cal P} \equiv \int \ du \ \vert u \ke \br u \vert . \eqno (1.10)
$$
We can show (see Appendix A)
that ${\cal P}$ has the property of a projection operator
${\cal P}^{2}={\cal P}$, if we use the appropriate Haar measure for
$SU(K)$ in $du$.

The partition function of the fermionic system in the subspace defined
by ${\cal P}$ may be expressed as a path integral, which is obtained
in the usual manner:
$$
Z = \int {\cal D}u e^{i\int dt\ \br u(t)\vert ( i\del_{t}-H ) \vert u(t) \ke }, 
\eqno (1.11)
$$
where ${\cal D}u$ is the appropriate Haar measure. The effective Lagrangian in 
the subspace defined by ${\cal P}$ is thus obtained as
$$
L_{eff} = \br u(t)\vert (i\del_{t}-H ) \vert u(t) \ke . \eqno (1.12)
$$
Recalling the definition of $\vert u \ke $ from $(1.8)$, we see that
$$
\br u(t) \vert i\del_{t} \vert u(t) \ke = {\rm tr}
(\rho_{0}u^{\dag }i\del_{t}u), \eqno (1.13)
$$
where
$$
(\rho_{0})_{\a \b } \equiv \br u \vert \ytdag_{\b } \yt_{\a } \vert u \ke .
\eqno (1.14)
$$
Thus,
$$
\eqalignno
{
(\rho_{0})_{\a \b } &= \d_{\a \b } , \a , \b \leq N \cr 
&= 0 \ \ {\rm otherwise} . & (1.15) \cr 
} 
$$
Let us consider a many body Hamiltonian of the form
$$ 
H = H^{(1)} + H^{(2)},
$$
with
$$
H^{(1)} = \y^{\dag }_{\a } h^{(1)}_{\a \b } \y_{\b } \eqno(1.16)
$$
and 
$$
H^{(2)} = \y^{\dag }_{\a_1}\ \y^{\dag }_{\a_2}\ h^{(2)\a_1 \a_2}_{\b_1 \b_2}\ 
\y_{\b_1}\ \y_{\b_2}. \eqno (1.17)
$$
The corresponding effective Hamiltonian is then given by
$$
H_{eff}\equiv \br u \vert H \vert u \ke = {\rm tr} \r_{0}u^{\dag }h^{(1)}u + 
\bigl( u\r_{0}u^{\dag }\bigr)_{\b_1 \a_1 }\ 
\bigl( u\r_{0}u^{\dag }\bigr)_{\b_2 \a_2 }\ \bigl[ h^{(2)\a_1 \a_2}_{\b_1 \b_2}
- h^{(2)\a_1 \a_2}_{\b_2 \b_1} \bigr]  \eqno (1.18)
$$
and the effective Lagrangian is, from $(1.13)$ and $(1.18)$, 
$$
\eqalignno{
L_{eff}  &=  {\rm tr}\ (\r_{0} u^{\dag }(i\del_{t} - h^{(1)}u)  - \cr 
&\sum_{\a_{1},\a_{2},\b_{1},\b_{2}}
\biggl( 
\bigl( u\r_{0}u^{\dag }\bigr)_{\b_1 \a_1 }\
\bigl( u\r_{0}u^{\dag }\bigr)_{\b_2 \a_2 }\ \bigl[ h^{(2)\a_1 \a_2}_{\b_1 \b_2}
- h^{(2)\a_1 \a_2}_{\b_2 \b_1} \bigr] \biggr) . & (1.19) \cr 
}
$$
The third term in $(1.19)$ is the so called Direct term and the last term, the
Exchange term.

We have thus obtained the effective Lagrangian governing the dynamics of the 
bosonic collective coordinate $u$ starting from the fermionic second 
quantised action. 
\bigskip
\centerline{\bf II. Planar Fermions in the Lowest Landau Level} 
\bigskip
Before introducing the bosonic collective coordinates, for the particular
case of the Hall ferromagnet, let us briefly 
recapitulate the basic physics of planar fermions subjected to a strong
magnetic field orthogonal to the plane.  We assume from the onset that
the Zeeman coupling of the electron spins to this magnetic field is zero,
due to the vanishingly small value of the effective g-factor.

The single particle Hamiltonian is the celebrated Landau Hamiltonian.
If $\vec A$ is the gauge potential (we choose the symmetric gauge for
convenience) giving rise to the external magnetic field, we have
$$
h_{0} = {{1}\over{2m}}\bigl( \vec p - \vec A \bigr)^{2} \equiv 
{1\over{2m}}{\vec \P }^{2} \eqno (2.1)
$$
where $\P^{x} = -i\del_{x} -{{B}\over2}y,\ \P^{y} = -i\del_{y} +{{B}\over2}x$.
Defining $\p \equiv {1\over{\sqrt{2B}}}\bigl( \P^{x} - i \P^{y} \bigr)$
and $\p^{\dag }$ as its complex conjugate, we have,
$$
\bigl[ \p , \p^{\dag } \bigr] = 1 .\eqno (2.2)
$$
The large degeneracy (${{B}\over{2\p }}$ states per unit area) of the
single particle spectrum is expressed through the introduction of the guiding centre
coordinates:
$$
\hat X\equiv \hat x-{1\over B}\hat \P^y ,\qquad\hat Y\equiv \hat y+{1\over B}
\hat \P^x , \ \ \ \ \ \ [\hat X,\hat Y]={i\over B}\ .\eqno (2.3)
$$
The holomorphic combination and its complex conjugate,
$$
\hat a\equiv \sqrt {B\over2}(\hat X+i\hat Y) , \ \ \ \ \hat
a^\dag \equiv \sqrt {B\over2}(\hat X-i\hat Y) \ ,\eqno (2.4)
$$
satisfy
$$[\hat \p ,\hat \p^\dag ]=1,\ \ \ \
[\hat a, \hat a^\dag ]=1,\ \ \ \
[\hat a, \hat \p ]=
[\hat a, \hat \p^\dag ]=[\hat a^\dag, \hat \p ]=
[\hat a^\dag, \hat \p^\dag ]=0\ .\eqno (2.5)
$$
These are related to the coordinate operators by
$$
{\sqrt{B\over 2}}(\hat x +i\hat y )\equiv \hat z =
\hat a -i \hat\p^{\dag} ,\ \ \
{\sqrt{B\over 2}}(\hat x -i\hat y )\equiv \hat{\zbar} =
\hat{ a}^{\dag} +i \hat\p \ .\eqno (2.6)
$$
Of course $[\hat z ,\hat{\zbar }]=0$ .

The eigenbasis of $h_0$ may be taken to be $\vert n,l \ke $, where
$n,l=0,1,2 \cdots \infty $. The index
$n$ is called the Landau level index. $n=0$ corresponds to the
lowest Landau level (L.L.L.). For a given $n$, $l$ measures the degeneracy
of the energy eigenstate.
We have
$$
\hat \p \vert n,l \ke = \sqrt{n} \vert n-1,l \ke ,\ \ \ \ \hat \p^{\dag } 
\vert n,l \ke 
= \sqrt{n+1} \vert n+1,l \ke \eqno (2.7)
$$
and 
$$
\hat a \vert n,l \ke = \sqrt{l} \vert n, l-1 \ke , \ \ \ \ \hat a^{\dag } 
\vert n,l \ke = \sqrt{l+1} \vert n,l+1 \ke . \eqno (2.8)
$$
Thus,
$$
\hat \p^{\dag } \hat \p \vert n,l \ke = n \vert n,l \ke ,\ \ \ \ 
\hat a^{\dag } \hat a \vert n,l \ke = l \vert n,l \ke . \eqno (2.9)
$$
We can also define a coherent state basis for $\hat a$. If
$$
\vert \x \ke \equiv e^{\x {\hat a}^{\dag }} \vert 0 \ke \eqno (2.10)
$$
we can easily check that $\hat a \vert \x \ke = \x \vert \x \ke $.
The inner product of two coherent states is given by 
$\br \j \vert \x \ke = e^{\bar \j \x } $ and the resolution of the
identity is $\int d^2 \x e^{-\vert \x \vert^2 } \vert \x \ke 
\br \x \vert = I $, where $d^2 \x \equiv {{d\ Re \x d\ Im \x }\over{\p }}
$.
The coherent state basis is related to the $\vert l \ke $ basis through
$\br l \vert \x \ke = {{\x^{l}}\over{\sqrt{l}}}$.
The L.L.L. wave function is given by 
$$
\br \vec x \vert 0,l\ke = \sqrt{{{B}\over{2\p }}}{1\over{\sqrt l}}
e^{-{1\over {2}}\vert z \vert^2 } {\bar z}^{l} \eqno (2.11)
$$
in the $\vert l \ke $ basis and
$$
\br \vec x \vert 0,\x \ke = \sqrt{{{B}\over{2\p }}}
e^{-{1\over 2}\vert z \vert^2 + \bar z \x } \eqno (2.12)
$$
in the coherent state basis.

A smooth function of $\hat z , \hat{\zbar }$, may be expanded in the
following manner:
$$
A(\hat z , \hat{\zbar }) = A(\hat a - i\hat{\p^{\dag }},\hat{a^{\dag }}+i
\hat p)=\sum_{p,q}{{(-i^{p})}{i^{q}}\over{p!\ q!}}(\hat{\p^{\dag }})^{p}
(\hat \p )^{q} \ddag \ \del^{p}_{z}\ \del^{q}_{\bar z}A(z,\bar z)\vert_
{z=\hat a , \bar z = \hat{a^{\dag }}}\  \ddag . \eqno (2.13)
$$
Here, the symbol $\ddag \cdots \ddag $ indicates that since the 
$\hat \p , \hat{\p^{\dag }}$ have been normal ordered, the 
$\hat a, \hat{a^{\dag }}$ are automatically {\it anti-normal ordered}.
Now when such a function is projected onto the L.L.L., only the term
with $p=0,q=0$ survives. Thus, in the L.L.L.,
the function $A(\hat z,\hat {\bar z})\rightarrow \ddag \ A(\hat a, \hat{a^{\dag }})
\ \ddag$. 
It is instructive to express this anti-normal ordered operator in the coherent 
state basis. Let
$$
\ddag \ A(\hat a, \hat{a^{\dag }}) \ \ddag 
\equiv \sum_{p,q} {1\over{p!\ q!}}(\hat a)^{p}\ 
(\hat{a^{\dag }}
)^{q}A_{pq}. \eqno (2.14)
$$
We now insert the resolution of the identity in the coherent state basis
between the $\hat a$s and the $\hat{a^{\dag }}$s in $(2.14)$.
We then obtain
$$
\ddag \ A(\hat a, \hat{a^{\dag }}) \ \ddag = \int d^2 \x e^{-\vert \x \vert^{2}}
\vert \x \ke A(\x ,\bar{\x })\br \bar{\x }
\vert \eqno (2.15)
$$
where $A(\x ,\bar{\x })\equiv \sum_{p,q} {1\over{p!\ q!}}
A_{pq} (\x )^{p}(\bar{\x })^{q}$.
Thus the product of two individually anti-normal ordered operators is
given by
$$
\ddag A(\hat a, \hat{a^{\dag }}) \ddag \ \ddag B(\hat a, \hat{a^{\dag }}) \ddag 
= \int d^2 \x e^{-\vert \x \vert^{2}} 
\vert \x \ke A(\x ,\bar{\x })
* B(\x ,\bar{\x })\br \bar{\x }
\vert \eqno (2.16)
$$
where the $* $ product is given by
$$
A(\x ,\bar{\x })* B(\x ,\bar{\x }) \equiv \sum^{\infty }_{n=0}
{{(-1)^{n}}\over{n!}}\del^{n}_{\bar \x }A\ \del^{n}_{\x }B . \eqno (2.17)
$$
The star product is associative in that $(A* \ B)* \ C = A*\ (B* \ C)$.
This concept of the star product should be very familiar to the aficionados of
non-commutative field theories. In fact, the field theory of fermions in the
L.L.L. is an instance of such a field theory, where the non-commutativity
is restricted to the spatial coordinates.

Upto this point, we have considered only the single particle Landau Hamiltonian.
However, there are other contributions to the many particle Hamiltonian,
which we shall now discuss.

The interaction between the electrons is the Coulomb interaction, whose
contribution, as indicated in $(1.18)$, splits naturally into a direct
part and an exchange part. It is the exchange part which is instrumental
in producing ferromagnetic behaviour in the Hall droplet.

Projected on to the L.L.L., the Coulomb term is written as:
$$
H_{c} = {1\over 2}\int d^2z_{1}\ d^{2}z_{2}\ e^{-\vert z_{1} \vert^{2}
-\vert z_{2} \vert^{2}}\ {\y }^\dag _{\a }(z_{1}){\y }^\dag _{\b }(z_{2})
V(\sqrt{{2\over{B}}}\vert z_{1}-z_{2} \vert )\y_{\b }(\bar{z}_{2})
\y_{\a }(\bar{z}_{1}) \eqno (2.18)
$$
where $V$ is the Coulomb interaction and $\y_{\a }$ are the second
quantised electron operators with spin $\a $, projected onto the L.L.L..
Let $\vert \y_{\a } \ke $ be the abstract
notation for the second quantised electron operator, with 
$\br \vec x \vert \y_{\a } \ke \equiv \y_{\a }(\vec x)$ being
the corresponding field operator. Then projection to the L.L.L.
entails:
$$
\br \vec x \vert \y_{\a } \ke \equiv \y_{\a }(\vec x) 
\rightarrow \br \vec x \vert \sum^{\infty }_{l=0}\vert 0,l \ke 
\br 0,l \vert \y_{\a } \ke . \eqno (2.19)
$$
We define $c_{\a }(l) \equiv \br 0,l \vert \y \ke $. This is the
operator that destroys an electron, in the L.L.L., with index $l$ and spin $\a $.
Then upon using $(2.11)$, we have 
$$
\y_{\a }(\vec x) \to \sqrt{{{B}\over{2\p }}}\sum^{\infty }_{l=0}
{1\over{\sqrt l}}
e^{-{1\over {2}}\vert z \vert^2 } {\bar z}^{l} c_{\a }(l) 
\equiv e^{-{1\over {2}}\vert z \vert^2 }\y_{\a }(\bar{z}).
\eqno (2.20)
$$
This is the $\y (\bar z) _\a$ that appears in  $(2.18)$.

Apart from the Coulomb interaction, we require that the electrons
be confined to a finite portion of the plane. This requires
the introduction of a suitable confining potential.

Let us introduce a radially symmetric confining potential $v(r)$ that
confines $N$ electrons in a droplet of radius $R$.
$$
v_{c}(\hat z, \hat{\zbar }) = \g \vert \hat{z} \vert^2 , \eqno (2.21)
$$
where $\g $ is the strength of the confining potential.

In view of the fact that the coordinate operators do not commute
when projected to the L.L.L.,the confining potential
is really the Hamiltonian of a one dimensional harmonic oscillator,
with the coordinate operators acting as canonically conjugate variables.
The eigenenergies of $v$ are therefore given by
$$
\e_{n} = \g (n+{1\over 2}), \ \ \ \ \ n=0,1,2, \cdots \infty . \eqno (2.22)
$$
If we fill the available single particle states with $N$ particles
to form a droplet with the lowest possible energy, the energy of
this droplet would be
$$
E_{tot}\equiv \sum^{N-1}_{n=0}\ \e_{n} = \g \ {{N^2}\over {2}}.
\eqno (2.23)
$$
We know that the degeneracy in the L.L.L. is given by ${{B}\over{2\p }}$
particles per unit area. If the radius of the droplet is $R$, we have
the relation
$$
N = \bigl( {{B}\over{2\p }}\bigr) \bigl( \p R^2 \bigr ) \Rightarrow 
N = {{B R^2}\over{2}} . \eqno (2.24)
$$
This tells us that if the strength of the magnetic field is held fixed
the area of the droplet scales with the number of electrons.
\footnote*{
From
$(2.23)$, it is clear that $E_{tot} \sim N^2$ as this is done. Thus,
from $(2.24)$, it seems that the energy increases quadratically
with the size of the sample. However, for the energy to be a
properly extensive quantity, we require that it should be directly
proportional to the area (i.e. to $N$). The way to resolve this is to
consider a
confining potential whose strength is of order one (in $N$) in the bulk
but is of order $N$ at the boundary. This form of the confining
potential shall be seen to have a crucial significance in determining
the leading order terms in the bosonic effective Lagrangian that we
shall compute further on.}

The ground state of the many electron system is formed by filling
up the single particle states of the confining potential sequentially with
electrons. Furthermore, as is well known, the ground
state of the system is ferromagnetic for large $N$. Thus the 
electrons in the ground
state all have say, spin ``up". Let us now try to relate all this to 
the formulation of the collective theory given in section I. The 
available sites ($K$ in number) correspond to the single particle eigenstates
of the confining potential (labelled by integers). Furthermore, each
state has two values of the spin index associated with it. Thus, in
line with what we have said in section I, the bosonic collective
fields are now expressed in terms of unitary operators belonging
to the fundamental representation of $SU(2K)$, with $K \to \infty $.
Similarly, the operator $\hat \r_{0}$ of section I is given in the present
context by
$$
\hat{\r_{0}} = \sum^{N-1}_{n=0} \vert n \ke \br n \vert \ \otimes \W \eqno (2.25)
$$
where $\W $ is a $2\times 2$ Hermitean matrix incorporating the
information about the spin of the many body ground state.
Since the ground state of the system
is ferromagnetic, $\W = P_{+} \equiv {1\over 2}(I+\s_{3})$.
$P_{+}$ is the projector onto the spin ``up" state.

We can express the operator $\hat \r_{0}$ in the coherent state basis:
$$
\r_{0}(\vert z \vert^{2}) \equiv e^{-\vert z \vert^{2}} \br \zbar 
\vert \hat \r_{0}(\hat a, \hat a^{\dag })\vert z \ke . \eqno (2.26)
$$
For a large number of particles 
($N \to \infty )$, 
$$
\r_{0}(\vert z \vert^{2}) = \theta (N-\vert z \vert^{2})\ P_{+}. 
\eqno (2.27)
$$
This was obtained in [9]. We have reproduced the derivation, for completeness,
in Appendix B.

$(2.27)$ immediately tells us that in order to convert this into a 
statement about the spatial extent of the system, we should require that
$\vert z \vert^{2} = {{N}\over{R^{2}}}r^{2}$, whence the theta
function would look like $\theta (N-{{N}\over{R^{2}}}r^{2})=
\theta (R^{2} - r^{2})$. This in turn tells us that  $z\sim \sqrt{N}$. 

\bigskip
\centerline{\bf III. Effective Lagrangian for the Collective Excitations
of the Quantum Hall Ferromagnet}
\bigskip
In this section, we shall present a somewhat detailed derivation of the
bosonic effective Lagrangian given in $(1.19)$, for the case of the 
quantum Hall ferromagnet. In the sequel, we shall take all the operators
to have been projected onto the L.L.L..

Since the operator $\hat u$ is unitary, we can write it as $\hat u 
\equiv e^{i \hat A}$, where $\hat A$ is a Hermitean operator. Let
us also define the function $g(z,\zbar ) \equiv e^{iA(z,\zbar )}$.
As we have noted in the previous section, $z \sim \sqrt{N}$. Thus
a derivative with respect to $z$, acting on $u(z,\zbar )$ will carry
with it a factor of $N^{-{1\over 2}}$. For a large value of $N$, we
thus have a natural parameter to expand $u$ in.
In fact,  
$$
u(z,\zbar ) = g(z,\zbar )\ - \ g_{,\zbar , z}\  + O({1\over{N^2}})
\eqno (3.1)
$$
where 
$g_{,\zbar , z}\ \equiv \ i\ g \ \int^{1}_{0} d \a \ e^{-i\a A}\ 
\del_{\zbar }A\ \del_{z}e^{i\a A}$.

Now, using the coherent state basis and the star product defined earlier,
we have
$$
\eqalignno{
{\cal A}\equiv 
{\rm tr} \ \hat{\r_{0}}\ {\hat u}^{\dag } \ i \del_{t}\hat u 
\ &= \ i\ {\rm tr}\ \int d^{2}z\ \r_{0}(\vert z \vert^{2})\ 
u(z,\zbar ,t)^{\dag }\ * \ \del_{t} u(z,\zbar ,t) \cr 
&= \ i\ {\rm tr}\ \int d^{2}z\ \q (N-\vert z \vert^2)\ P_{+}\
u(z,\zbar ,t)^{\dag }\ * \ \del_{t} u(z,\zbar ,t), & (3.2) \cr 
}
$$
where the trace on the r.h.s. is over the spin indices.
Using $(3.1)$, we obtain,
$$
{\cal A} =  
i \int d^{2}z\ \q (N-\vert z \vert^2)\ {\rm tr}P_{+}\ 
\bigl[\ g^{\dag }\ \del_{t}g\ - \ \del_{\zbar }g^{\dag }\ \del_{z}
\del_{t}g\ - \ g^{\dag }\ \del_{t}g_{,\zbar , z}\ - \ (g_{,\zbar , z})^
{\dag }\ \del_{t}g\ +\ \cdots \bigr] . \eqno (3.3)
$$
After some rather straightforward manipulation of the above equation,
using the relation $g^{\dag }\ g_{,\zbar , z}\ +\ c.c.\ =\ 
g^{\dag }\ \del_{\zbar }g\ g^{\dag }\ \del_{z}g$, 
we get
$$
{\cal A} = {\cal A}_{bulk}\ +\ {\cal A}_{edge}, \eqno (3.4)
$$
where
$$
{\cal A}_{bulk}\ \equiv \ 
i \int d^{2}z\ \q (N-\vert z \vert^2)\ {\rm tr}P_{+}\ g^{\dag }\ 
\del_{t}g \eqno (3.5)
$$
and
$$
{\cal A}_{edge}\ \equiv \ 
{{i}\over{2}} \int d^{2}z\ \d (N-\vert z \vert^2)\ {\rm tr}P_{+}\ 
\biggl( g^{\dag }\ (\zbar\ \del_{\zbar }\ - \ z\ \del_{z} )\ g\ 
g^{\dag }\ \del_{t}g\ +\ [g^{\dag }\ z\del_{z}g,\ g^{\dag }\ \del_{t}g ]
\biggr) . \eqno (3.6)
$$
Equation $(3.5)$ is a leading contribution to the effective
Lagrangian for the bulk and $(3.6)$, a leading contribution to
the edge dynamics.

The second term in $(1.19)$ is the contribution of the confining potential
$v_{c}$ to the effective Lagrangian. In the coherent state representation,
it is written as:
$$
{\cal B}\equiv -{\rm tr}\ \hat{\r_{0}}\ \hat{u^{\dag }}\ \hat{v_{c}}\ \hat{u}\ = \ 
-\ {\rm tr}\ \int d^{2}z\ \q (N-\vert z \vert^2)\ P_{+}\ 
(u^{\dag }\ *\ v_{c}\ *\ u) . \eqno (3.7)
$$
Using $(3.1)$ and the definition 
of the star product, we get, after a simple calculation,
$$
\eqalignno{
{\cal B}&=\int d^2z \ \theta (N-\vert z \vert^2) [-v_{c} + {\rm tr}\ P_{+}
(\del_{z}v_{c}g^{\dag }\del_{\bar z}g - \del_{\bar z}v_{c}g^{\dag }\del_{z}g) ]\cr 
&+\int d^{2}z \ \delta(N-\vert z \vert^2)v_{c}{\rm tr}\ P_{+}g^{\dag }\del_{\bar
 z}g
\ g^{\dag }\del_{z}g . & (3.8) \cr
}
$$
The first term in $(3.8)$ is the bulk term and the second, the edge
contribution.

Let us now look at the contribution of the exchange part of the 
Coulomb interaction to the effective Lagrangian.
In equation $(2.18)$, we have already given the form of the Coulomb
interaction, projected to the L.L.L.. We shall start from this expression 
and obtain the exchange part from it.
Before doing so, let us establish some notation.
Let $k\equiv {1\over{\sqrt{2B}}} (k_{x} + i k_{y})$ and $\bar{k}$ be
its complex conjugate. Let $V(k,\bar{k})$ be the Fourier transform of
the Coulomb interaction. In this notation,
$$
V(\vert \vec x_{1} - \vec x_{2} \vert ) = 
\int d\vec k \ V(k,\bar{k})\ 
e^{-i(k\bar{z_{1}}+\bar{k}z_{1})}\ e^{i(k\bar{z_{2}}+\bar{k}z_{2})}. 
\eqno (3.9) 
$$
Further, adopting the notation introduced in equation $(2.19)$, we
obtain
$$
H_{C}\ =\ 
{1\over 2}\int d \vec k \ V(k,\bar{k})\  \br \psi_{\a } \vert \ 
e^{-i{\hat \chi }}\ \vert \psi_{\a } \ke \ 
\br \psi_{\b } \vert \ e^{i{\hat \chi }}\ \vert \psi_{\b } \ke 
\eqno (3.10) 
$$
where $e^{i{\hat \chi }}\equiv e^{i\bar{k} \hat{a}}\ 
e^{ik \hat{a}^{\dag }}$ and $e^{-i{\hat \chi }}$ is its Hermitean
conjugate.
Let us further denote $e^{i\chi (1) }\equiv e^{i\bar{k} z_{1}}\ 
e^{ik \zbar_{1}}$ and similarly for $e^{i\chi (2)}$.

In the notation of $(1.19)$, 
$$
h^{(2)\b_{1} \b_{2}}_{\a_{1} \a_{2}} = {1\over 2}\int d \vec k \ 
V(k,\bar{k})
\ \bigl( e^{-i\hat{\chi }} \bigr)_{\a_{1} \b_{1}}\ 
\bigl( e^{i\hat{\chi }} \bigr)_{\a_{2} \b_{2}}. \eqno (3.11)
$$
Hence, the expectation value of the Coulomb Hamiltonian in the state
$\vert u \ke $ is
$$
\br H_{C} \ke = {1\over 2}\int d \vec k \ V(k,\bar{k})\ (\hat{u} 
\hat{\r_{0}} {\hat u}^{\dag })_{\b_{1}\a_{1}}\ 
(\hat{u} 
\hat{\r_{0}} {\hat u}^{\dag })_{\b_{2}\a_{2}}\ 
\biggl[ \bigl( e^{-i\hat{\chi } } \bigr)_{\a_{1}\b_{1}}\ 
\bigl( e^{i\hat{\chi } } \bigr)_{\a_{2}\b_{2}}\ - \ 
\bigl( e^{-i\hat{\chi } } \bigr)_{\a_{1}\b_{2}}\ 
\bigl( e^{i\hat{\chi } } \bigr)_{\a_{2}\b_{1}} \biggr] . \eqno (3.12)
$$ 
The first term is the direct contribution of the Coulomb term and the 
second, the exchange contribution.

Let us first focus on the exchange contribution.
>From $(3.12)$, it is given by
$$
{\cal C} \equiv L^{(ex)}_{eff} = {1\over 2}\int d \vec k \ V(k,\bar{k})\ 
{\rm tr}\ \bigl( \hat{\r_{0}}{\hat u}^{\dag }e^{-i\hat{\chi } }{\hat u}\ 
\hat{\r_{0}}{\hat u}^{\dag }e^{i\hat{\chi } }{\hat u} \bigr) . \eqno (3.13)
$$
Then, introducing the coherent state basis and the star product, we
write the contribution to the effective bosonic Lagrangian, of
the exchange part as 
$$
\eqalignno{
{\cal C} = {1\over 2}\int d \vec k \ V(k,\bar{k})
&\int d^{2}z_{1}\ d^{2}z_{2}e^{-\vert z_{1} \vert^{2} }
e^{-\vert z_{2} \vert^{2}}
{\rm tr}\biggl[ \hat{\r_{0}}\ \vert z_{1} \ke 
(u^{\dag }*e^{-i\chi (1)}*u)(z_{1},\zbar_{1}) \cr 
& \br \bar{z_{1}}\vert \hat{\r_{0}}\vert z_{2}\ke 
(u^{\dag }*e^{i\chi (2)}*u)(z_{2},\zbar_{2})\br \bar{z_{2}} \vert \biggr] 
. & (3.14)
\cr}
$$
Thus,
$$
\eqalignno{
{\cal C} = {1\over 2}\int d \vec k \ &V(k,\bar{k})
\int d^{2}z_{1}\ d^{2}z_{2}e^{-\vert z_{1} \vert^{2} }
e^{-\vert z_{2} \vert^{2}}
\vert \br \bar{z_{1}} \vert \hat{\r_{0}} \vert z_{2} \ke \vert^{2} \cr 
&{\rm tr}P_{+}(u^{\dag }*e^{-i\chi (1)}*u)(z_{1},\zbar_{1})
P_{+}(u^{\dag }*e^{i\chi (2)}*u)(z_{2},\zbar_{2}). & (3.15)
\cr} 
$$
The spin projector, $P_{+}$, has been explicitly written in $(3.15)$.


After a somewhat lengthy calculation, presented in detail in Appendix C
and D, we obtain
$$
\eqalignno{
{\cal C}= &- {{1}\over{2l}}\sqrt{{{\pi }\over{2}}}\int d^2z\ 
\theta (N-\vert z \vert^{2})
[{\rm tr}\ P_{+}g^{\dag }\del_{\bar z}g g^{\dag }\del_{z}g - {\rm tr}\ P_{+}
g^{\dag }\del_{\bar z}g
{\rm tr}\ P_{+}g^{\dag }\del_{z}g] \cr
&+{{1}\over{4l}}\sqrt{{{\pi }\over{2}}}\int d^2z \ \delta(N-\vert z \vert^2)
{\rm tr}\ P_{+}g^{\dag }(\bar z \del_{\bar z}-z\del_{z})g . & (3.16) \cr
}
$$
Here $l={{1}\over{\sqrt{B}}}$ is the magnetic length.
The first term is the bulk contribution and the second, that from the edge.

The contribution of the direct term is 
given by (see $(3.12)$),
$$
{\cal D}\equiv L^{(dir)}_{eff}
= -{1\over {2}}\int d^{2}k\ V(k,\bar{k})\ 
{\rm tr}\bigl( \hat{\r_{0}} {\hat{u}}^{\dag } e^{-i\hat{\chi }} \hat{u}
\bigr) \ 
{\rm tr}\bigl( \hat{\r_{0}} {\hat{u}}^{\dag } e^{i\hat{\chi }} \hat{u}
\bigr) . \eqno (3.17)
$$
Upon introducing the coherent states, we obtain
$$
\eqalignno {{\cal D}
= -{1\over {2}}\int d^{2}k\ V(k,\bar{k})\
&\int d^{2}z_{1} d^{2}z_{2} \ 
\q(N-\vert z_{1} \vert^{2})\ 
\q(N-\vert z_{2} \vert^{2})\cr 
&\ {\rm tr}P_{+}\bigl( u^{\dag }*e^{-i\chi (1)}*
u \bigr) \ {\rm tr}P_{+}\bigl( u^{\dag }*e^{i\chi (2)}*
u \bigr) . & (3.18)\cr }
$$
Now, by using the definition of the star product and the expansion of
$u$ in terms of $g$, we may, upon suitable integration by parts,
obtain
$$
\eqalignno {
\int d^{2}z_{1}\ \q(N-\vert z_{1} \vert^{2})\ 
{\rm tr}P_{+}\bigl( u^{\dag }&*e^{-i\chi (1)}*
u \bigr) \simeq 
\int d^{2}z_{1}\ \q(N-\vert z_{1} \vert^{2})\ 
{\rm tr}P_{+}\bigl( 1 - [g^{\dag }\del_{\zbar }g,g^{\dag }\del_{z}g]
\bigr) e^{-i\chi (1)} \cr 
&+\int d^{2}z_{1}\ \d(N-\vert z_{1} \vert^{2})\ {\rm tr}P_{+}
g^{\dag }(\zbar \del_{\zbar }-z\del_{z})g e^{-i\chi (1)} .  
& (3.19)\cr 
}
$$
We drop the term that does not involve $g$ from $(3.20)$ as it is
the Coulomb contribution to the fermionic ground state and not to
the collective excitations. Further, we can easily convince ourselves
that $\d \r (z,\zbar )$, which is the deviation of the mean local density 
from its constant ground state value, is given by
$$
\d \r (z,\zbar )=
-\q (N-\vert z \vert^{2})\ {\rm tr}P_{+}
[g^{\dag }\del_{\zbar }g,g^{\dag }\del_{z}g]
\ + \ \d(N-\vert z_{1} \vert^{2})\ {\rm tr}P_{+}
g^{\dag }(\zbar \del_{\zbar }-z\del_{z})g . \eqno (3.20)
$$
Thus, using $(3.19)$ and $(3.20)$, we can rewrite $(3.18)$ as
$$
{\cal D} = -{1\over {2}}\int d\vec x_{1}d\vec x_{2}
\ V(\vert \vec x_{1}
-\vec x_{2} \vert )\ \d \r (\vec x_{1})\ \d \r (\vec x_{1}). \eqno (3.21)
$$
Interestingly enough, the deviation from the mean ground state density
picks up contributions both in the bulk and at the edge.
\bigskip
\centerline{\bf IV. Scaling of the terms in the bosonic effective 
Lagrangian with $N$} 
\bigskip
In the preceding section, we have obtained the leading contributions
to the bosonic effective Lagrangian that emerge from the underlying
microscopic fermionic action. We have considered the number of 
particles in the system, $N \gg 1$ and have developed a systematic 
derivative expansion scheme, with $1/\sqrt{N}$ as the small parameter, to
identify the leading contributions. In this section, we shall study
how the various terms scale with $N$ and rewrite the various terms
as integrals over real spatial coordinates.
In equation $(3.5)$, we note that in order to write the theta function
in terms of spatial coordinates, we have to take $z = {{\sqrt{N}}\over{R}}
\j $, where $R$ is the radius of the droplet given by $R=\sqrt{{{2N}\over
{B}}}$ and $\j \equiv x+iy$.
Then we get $\q(N-\vert z \vert^{2}) = \q(R^{2}-r^{2})$. This
just means that the bulk contribution has support inside of the droplet.
The measure $d^{2}z$ then becomes ${{B}\over{2\p }}dx\ dy$.
Thus, we have, from $(3.5)$,
$$
{\cal A}_{bulk}
=
i{{B}\over{2\p }}\int d\vec x \ \q(R^{2}-r^{2})\ {\rm tr} 
P_{+}\ g^{\dag }\ \del_{t}g . \eqno (4.1)
$$
This term is proportional to the area of the droplet and hence is
proportional to $N$.
Again, from $(3.6)$, we have,
$$
{\cal A}_{edge} =
{{1}\over{8\p }}\int^{2\p }_{0} d\q \ {\rm tr}P_{+}\
\bigl( \{ g^{\dag }i\del_{\q }g,g^{\dag }i\del_{t}g \} \ + \ [g^{\dag }r\del_{r}
g,g^{\dag }i\del_{t}g]
\bigr)_{r=R} . \eqno (4.2)
$$
This term is of $O(1)$ in $N$ as expected, as it is a boundary term
and as such should be independent of the number of particles in the
droplet.

The contributions due to the confining potential are given in
$(3.8)$.
We have argued previously that for a proper thermodynamic limit to
exist, that is, for the energy of the droplet to scale as $N$, the
confining potential should be of order one (in $N$) in the bulk.
Alternatively, $\del_{z}v \sim 1/\sqrt{N}$ in the bulk. Thus, we
see that the second term in ${\cal B}$ is an order one contribution to the bulk
effective
Lagrangian. Thus it is a subleading contribution (compared to O($N$)
contributions) and may be dropped.
Again,
$$
{\cal B}_{edge}
=
-{{\w }\over{4\p }}\int^{2\p }_{0} d\q \ {\rm tr}P_{+}
\bigl(- (g^{\dag }r\del_{r}g)^{2} + [g^{\dag }r\del_{r}g,
g^{\dag }i\del_{\q }g] + (g^{\dag }i\del_{\q }g)^{2} \bigr)_{r=R}
\eqno (4.3)
$$
where $\w \equiv {{v(r=R)}\over {2N}}$, and is of order one in $N$.

Let us now turn our attention to the contributions of the Coulomb term.
$$
{\cal C}_{bulk}
=
-{1\over{2l}}{1\over{\sqrt{2\p }}}
\int d\vec x \ \q(R^{2}-r^{2})\bigl[
{\rm tr}P_{+}\ g^{\dag }\del_{\bar{\j }}g\ g^{\dag }\del_{\j }g\ - \
{\rm tr}P_{+}\ g^{\dag }\del_{\bar{\j }}g\
{\rm tr}P_{+}\ g^{\dag }\del_{\j }g \bigr] . \eqno (4.4)
$$
Similarly, using $\bar{\j }\del_{\bar{\j }}-\j \del_{\j }
=i\del_{\q }$, where $\q $ is the plane polar angle, we get
$$
{\cal C}_{edge}
=
{1\over{8l}}{1\over{\sqrt{2\p }}}\int^{2\p }_{0}d\q
\ {\rm tr}P_{+}\ g^{\dag }\ i\del_{\q }\ g \vert_{r=R}. \eqno (4.5)
$$
The contribution in $(4.4)$ is proportional to the area of the
droplet and is thus of order $N$, whilst that in $(4.5)$ is of
order one in $N$.

Similar arguments can be provided for the contributions from the direct
part of the Coulomb interaction.

The the leading contributions to the effective Lagrangian, governing
the collective excitations in the bulk, are given by
$$
\eqalignno
{
&L^{(bulk)}_{eff}=
\int_{{\cal D}}d\vec x \biggl[ {{B}\over{2\p }} {\rm tr}P_{+}
g^{\dag } i\del_{t}g -{1\over{2l}}{1\over{\sqrt{2\p }}}
\bigl( {\rm tr}P_{+}g^{\dag }\del_{\bar{\j }}g\ g^{\dag }\del_{\j }g
\ - \ {\rm tr}P_{+}g^{\dag }\del_{\bar{\j }}g\
{\rm tr}P_{+}g^{\dag }\del_{\j }g \bigr) \biggl] \cr
&
+{1\over{2{\p }^{2}}}\int_{{\cal D}}d\vec x_{1}\int_{{\cal D}}d\vec x_{2}\
V(\vert \vec x_{1}-\vec x_{2} \vert )\
\bigl( {\rm tr}P_{+}[g^{\dag }\del_{\bar{\j_{1}}}g,
g^{\dag }\del_{\j_{1}}g]\bigr )\
\bigl( {\rm tr}P_{+}[g^{\dag }\del_{\bar{\j_{2}}}g,
g^{\dag }\del_{\j_{2}}g]\bigr ) & (4.6) \cr
}
$$
The subscript ${\cal D}$ indicates that the integral is
over all $r < R$.

Similarly, we obtain the effective Lagrangian governing the collective
dynamics at the edge of the droplet
$$
\eqalignno
{
L^{(edge)}_{eff}=
&\int^{2\p }_{0}d\q \ \biggl[ {{1}\over{8\p }}
{\rm tr}P_{+}\
\bigl( \{ g^{\dag }i\del_{\q }g,g^{\dag }i\del_{t}g \} \ +
[g^{\dag }r\del_{r}g,g^{\dag }i\del_{t}g ]
\bigr)_{r=R}\cr
&-{{\w }\over{4\p }}
{\rm tr}P_{+}
\bigl(- (g^{\dag }r\del_{r}g)^{2} + [g^{\dag }r\del_{r}g,
g^{\dag }i\del_{\q }g] + (g^{\dag }i\del_{\q }g)^{2} \bigr)_{r=R}\cr
&-{1\over{2l}}{1\over{\sqrt{2\p }}}
{\rm tr}P_{+}\ g^{\dag }\ i\del_{\q }\ g \vert_{r=R}\biggl] \cr
&
-{1\over{8\p }}\int^{2\p }_{0}d\q_{1}d\q_{2}\ V(\vert \vec x_{1}
-\vec x_{2} \vert )\ ({\rm tr}P_{+}g^{\dag }i\del_{\q_{1}}g)
({\rm tr}P_{+}g^{\dag }i\del_{\q_{2}}g)\vert_{r_{1},r_{2}
=R} .
 & (4.7) \cr
}
$$

At this point, we note that if $\hat{\r_{0}}=\sum^{\infty }_{n=0}
\vert n \ke \br n \vert \ P_{+}$, all the available single particle states
in the L.L.L. would have been filled and the effective Lagrangian
would have been entirely a bulk effective Lagrangian
(as the droplet would
in this case fill the entire plane).
In fact, in this case,
$$\r_{0}(\vert z \vert^{2}) \equiv e^{-\vert z \vert^{2}}
\br \zbar \vert \hat{\r_{0}} \vert z \ke = P_{+}. \eqno (4.8)
$$
Thus, the effective Lagrangian would be given by $(4.6)$ with
the support of the integral over $x,y$ extending over the entire
droplet.
Namely,
$$
\eqalignno
{
&L^{(bulk)}_{eff}=
\int d\vec x \biggl[ {{B}\over{2\p }} {\rm tr}P_{+}
g^{\dag } i\del_{t}g -{1\over{2l}}{1\over{\sqrt{2\p }}}
\bigl( {\rm tr}P_{+}g^{\dag }\del_{\bar{\j }}g\ g^{\dag }\del_{\j }g
\ - \ {\rm tr}P_{+}g^{\dag }\del_{\bar{\j }}g\
{\rm tr}P_{+}g^{\dag }\del_{\j }g \bigr) \biggl] \cr
&
+{1\over{2{\p }^{2}}}\int_{{\cal D}}d\vec x_{1}\int_{{\cal D}}d\vec x_{2}\
v_{c}(\vert \vec x_{1}-\vec x_{2} \vert )\
\bigl( {\rm tr}P_{+}[g^{\dag }\del_{\bar{\j_{1}}}g,
g^{\dag }\del_{\j_{1}}g]\bigr )\
\bigl( {\rm tr}P_{+}[g^{\dag }\del_{\bar{\j_{2}}}g,
g^{\dag }\del_{\j_{2}}g]\bigr ) & (4.9) \cr
}
$$
This is the effective Lagrangian that has been discussed extensively
in the literature [1],[2],[5], in the context of ferromagnetic magnons in the
Hall ferromagnet.
\bigskip
\centerline{\bf V. Simplifying the effective Lagrangian}
\bigskip
In this section, we shall look closely at the various contributions to the
bulk and the edge effective Lagrangians and comment on their relative
importance. We shall look separately at the the bulk and the edge
contributions.

In terms of the familiar Euler angles, we can parametrise $g$ as
$$
g\equiv e^{-i{{\phi }\over{2}}\sigma_{3}}\ e^{-i{{\theta }\over{2}}\sigma_{2}}\
e^{-i{{\chi }\over{2}}\sigma_{3}}
\eqno (5.1)
$$
with
$$
g\sigma_{3}g^{\dag } \equiv \hat m \cdot \vec \sigma \ , \ {\hat m}^{2} = 1 \eqno
 (5.2)
$$
where
$$
\hat m = (\sin \theta \cos \phi , \sin \theta \sin \phi , \cos \theta ) .
$$

Alternatively, we may parametrise $g$ in a more conventional manner as
$$
\eqalignno{
g & \equiv e^{i\vec A \cdot \vec \sigma } \cr
g^{\dag }i\del_{\mu }g & = -\vec \sigma \cdot (\del_{\mu }\vec A + \vec A \times
 \del_{\mu }
\vec A ) . & (5.3) \cr
}
$$
where we have retained upto the quadratic in $\vec A$. Furthermore, we shall
assume the following
ansatz for $\vec A$ [2],
$$
\vec A = {{1}\over {2}}(m_{2}, -m_{1}, 0) \eqno (5.4)
$$
and consider $m_{3} \simeq 1$, with $\vert \vec m_{T} \vert \ll 1$, where $\vec
m_{T} \equiv
(m_{1},m_{2},0)$.

Then the various contributions to the bulk effective Lagrangian may be written as:

$$
\eqalignno{
{\cal A}_{bulk} & = -{{1}\over{2}}{{B}\over{2\pi }}
\int_{D} d\vec x (1-\cos \theta )\del_{t} \phi \cr 
& = {{1}\over{2}}{{B}\over{2\pi }}\int_{D} d\vec x \int^{1}_{0} d\lambda \  
{\hat m}_{\lambda }
(\vec x,t) \cdot [\del_{t}{\hat m}_{\lambda }(\vec x,t) \times 
\del_{\lambda }
{\hat m}_{\lambda }(\vec x,t)] & (5.5) \cr 
}
$$
where ${\hat m}_{\lambda } \equiv {\hat m}(\lambda \theta ,\phi )$.
We require ${\hat m}_{\lambda =0}={\hat e}_{3}$ and ${\hat m}_{\lambda =1}={\hat
 m}$. Explicitly,
we may choose
$$ {\hat m}_{\lambda }(\theta ,\phi ) = \bigl( \sin \lambda \theta \sin \phi ,
\sin \lambda \theta \cos \phi , \cos \lambda \theta \bigr) .$$
The second line in $(5.5)$ is the well-known
geometric phase generic to quantum ferromagnets 
and is proportional to the area of the droplet. The suffix $D$
indicates
that the integral has support over the entire area of the droplet.

As we have argued before, we may drop ${\cal B}_{bulk}$.

$$
{\cal C}_{bulk} = -{1\over{32l\sqrt{2\pi }}}\int_{D}d\vec x (\del_{\alpha }{\hat
 m})^{2}
-{{1}\over{4l}}\sqrt{{{\pi }\over{2}}}\int_{D}d\vec x \rho_{p}(\vec x,t) \eqno (
5.6)
$$
where $\rho_{p} \equiv -{1\over{8\pi }}\epsilon_{\alpha \beta }{\hat m}\cdot
(\del_{\alpha }{\hat m} \times \del_{\beta }{\hat m})$ is the well known 
Pontryagin index
density. The integral of $\rho_{p}$ over all space gives an integer, the
Pontryagin index, which is a topological quantity.

$$
{\cal D}_{bulk} = -{1\over{2}}\int_{D} d\vec x_{1} d\vec x_{2} \rho_{p}(\vec x_{
1},t)
V(\vert \vec x_{1} - \vec x_{2} \vert )\rho_{p}(\vec x_{2},t) \eqno (5.7).
$$

Thus, the bulk effective Lagrangian is given by:
$$
\eqalignno{
L^{(bulk)}_{eff} & = \int_{D} d\vec x \bigl[ {{B}\over{4\pi }} \int^{1}_{0}d
\lambda
{\hat m}_{\lambda }\cdot [\del_{t}{\hat m}_{\lambda }\times \del_{\lambda }{\hat
 m}_{\lambda }]
-{1\over{32l\sqrt{2\pi }}}(\del_{\alpha }{\hat m})^{2}-{{1}\over{4l}}\sqrt{{{\pi
 }\over{2}}}\rho_{p}
\bigr] \cr
&-{1\over{2}}\int_{D}d\vec x_{1}d\vec x_{2} \rho_{p}(\vec x_{1})V(\vert \vec x_{
1} - \vec x_{2} \vert )
\rho_{p}(\vec x_{2}) . & (5.8) \cr
}
$$

This Lagrangian has been obtained in a variety of ways in the literature. 
It governs the dynamics
of the ferromagnetic magnons [1],[2],[5].

Let us now look at the effective Lagrangian governing the edge excitations.
To simplify matters,
we shall focus on those excitations that satisfy the condition $\del_{r}{\hat m}
(r,\theta ,t)
\vert_{r=R} = 0$.

Then, using $(5.3),(5.4)$,
$$
{\cal A}_{edge} \simeq {1\over{16\pi }}\int^{2\pi }_{0}d\theta [\del_{t}{\hat m}
 \cdot
\del_{\theta }{\hat m} ]_{r=R} . \eqno (5.9)
$$
Further,
$$
{\cal B}_{edge} \simeq -{{\omega }\over{16 \pi }}\int^{2\pi }_{0}d\theta [(\del_
{\theta }
{\hat m} )^{2} ]_{r=R} \eqno (5.10)
$$
where $\omega \equiv {{v_{c}(r=R)}\over{2N}}$.

Interestingly enough, {\it to the leading order}, the contributions of the 
Coulomb interaction
to the Lagrangian for the edge, are zero.

Thus,
$$
L^{(edge)}_{eff} = {1\over{16 \pi }}\int^{2\pi }_{0}d\theta [\del_{t}{\hat m}
\cdot \del_{\theta }{\hat m} - {{\omega }}(\del_{\theta }{\hat m} )^{2}]
_{r=R} . \eqno (5.11)
$$

These edge excitations are obviously chiral in nature.

We see that even though the bulk modes (ferromagnetic magnons) and the edge modes
(chiral excitations)
are both gapless, they owe their dynamics to completely different sources. 
For the bulk modes,
which are the Goldstone modes corresponding to a spontaneous breaking of 
the global spin symmetry
$SU(2) \rightarrow u(1)$, the exchange part of the Coulomb interaction is 
not merely crucial to their
dynamics, but is truly their {\it raison d'etre}.

On the other hand, the chiral edge excitations are somewhat generic to confined
Hall fluids.
Let us consider a specific instance. We shall consider a conventional $\nu =1$ H
all droplet
where the spin degrees of freedom are completely frozen by the Zeeman term.
In this case,
$\rho(\vert z \vert^{2}) = \theta (N-\vert z \vert^{2})$ and $g$ is simply a $u(
1)$ phase.
The edge contribution to to the effective Lagrangian can be easily shown to be
$$
L^{(edge)}_{eff}={1\over{2 \pi }}\int^{2\pi }_{0}d\theta [\del_{\theta }\phi 
\del_{t}\phi  -
{{\omega }}(\del_{\theta }\phi )^{2}] , \eqno (5.12)
$$
where $g = e^{i\phi }$.

This is the familiar Lagrangian obtained in the literature [10] for a case where
the Coulomb interaction
is known to be unimportant. The corresponding bulk contribution trivially vanishes.

This should convince us that the edge excitations are extremely ubiquitous and 
as such do not owe
their existence to the Coulomb interaction. Thus, to the leading order, the
edge excitations of a quantum Hall ferromagnet are qualitatively similar 
to those for the standard $\nu =1$ integer Hall droplet.
\bigskip
\centerline{\bf VI. Conclusions}
\bigskip
In this article, we have utilised the bosonisation technique introduced in [8] 
to discuss the
collective excitations of a finite-sized Hall sample where the Land\'e 
$g$ factor is vanishingly small. 
For a large but finite sample,
(Area $ \gg {{1}\over{B}}$), we know that these excitations are 
ferromagnetic magnons
whose dynamics is governed by the exchange part of the Coulomb interaction 
between the electrons.

We have shown that the effective Lagrangian for the collective excitations  splits naturally
into two pieces, one having support in the bulk of the droplet and the 
other at the edge.

The bulk effective Lagrangian coincides with that computed for an infinite 
sample [1],[2],[5].
The edge excitations, which are chiral in nature, are to the leading order, 
unaffected by the
Coulomb interaction between the electrons. The mere fact that the electrons are
confined to
a droplet suffices to produce chiral edge excitations, which are 
qualitatively similar to those for the conventional $\nu = 1$ Hall
droplet.
\bigskip
\centerline{\bf Acknowledgements}
\bigskip
R.R. acknowledges discussions with D. Karabali, S. Hebboul, S. Mitra and 
R. Narayanan during the course of this work. 
\bigskip
\centerline{\bf References}
\bigskip
\item{[1]} S.L. Sondhi et. al. Phys. Rev. {\bf B 47}, 16419, (1993).
\item{[2]} K. Moon et. al. Phys. Rev. {\bf B 51}, 5138, (1995).
\item{[3]} S.M. Girvin, {\it The Quantum Hall Effect: Novel Excitations
and Broken Symmetries}, (Lectures delivered at Les Houches, 1997), for
an extensive review and bibliography. 
\item{[4]} S.E. Barrett et. al. , Phys. Rev. Lett. {\bf 74}, 5112, (1995);
A. Schmeller et. al. , Phys. Rev. Lett. {\bf 75}, 4290, (1995); 
E.H. Aifer, B.B. Goldberg and D.A. Broido, Phys. Rev. Lett. {\bf 76}, 680, (1996).
\item{[5]} R. Ray, Phys. Rev. {\bf B60}, 14154, (2000).
\item{[6]} X.G. Wen, Phys. Rev. {\bf B 41}, 12838, (1990); 
M. Stone, Ann. Phys. (N.Y.), {\bf 207}, 38, (1991).
\item{[7]} J.H. Oaknin, L. Martin-Moreno, Phys. Rev. {\bf B 54}, 16850, (1996); 
A. Karlhede et. al.,  Phys. Rev. Lett. {\bf 77}, 2061, (1996); 
K. Lejnell, A. Karlhede and S.L. Sondhi, Phys. Rev. {\bf B 59}, 10183, (1999); 
A. Karlhede, K. Lejnell and S.L. Sondhi, Phys. Rev. {\bf B 60}, 15948, (1999).
\item{[8]} B. Sakita, Phys. Lett. {\bf B 387}, 118, (1996) and references 
therein. 
In this paper 
the Lagrangian (1.19) was derived
by a gauge fixing procedure in the gauge theory description of collective coordinates,
without deeply questioning the implementability of this gauge fixing.
This gauge choice is not always possible. The
section I of the present paper replaces section 2 of [8].
\item{[9]} B. Sakita, Phys. Lett. {\bf B 315}, 124, (1993).
\item{[10]} S. Iso, D. Karabali and B. Sakita, Nucl. Phys. {\bf B 388}, 700, (1992)
; Phys. Lett. {\bf B 296}, 143, (1992). 
\vfill\eject
\bigskip
\centerline{\bf Appendix A}
\bigskip
Let ${\cal D}(u)$ denote the totally antisymmetric irreducible representation
of $SU(K)$, of dimensionality $K\choose{N}$.
Then
$$
{\cal D}_{A0}\ =\ {1\over{\sqrt{N!}}}\ \e_{\b_{1} \cdots \b_{N}}\ 
u_{\a_{1}\b_{1}}\cdots u_{\a_{N}\b_{N}} \eqno (A.1)
$$
where the suffix $0$ refers to the sequence $1,1,\cdots N$ and
$A$ runs over $K\choose{N}$ values.
Further, let 
$$\vert A \ke \ \equiv \ \vert \a_{1},\a_{2}\cdots \a_{N} \ke \eqno (A.2)
$$
where $\vert \a_{1},\a_{2}\cdots \a_{N} \ke $ has been given in $(1.3)$.
Thus, from $(1.3),(1.6),(A.1)$ and $(A.2)$, we get
$$
\vert u \ke \ = \ \sum_{A} \vert A \ke {\cal D}_{A0}(u) \eqno (A.3)
$$
and 
$$
\br u \vert \ = \ \sum_{A} \br A \vert {\cal D}^{\ast }_{A0}(u). \eqno (A.4)
$$
We further know that
$$
\int du\ {\cal D}_{A_{1}B_{1}}(u)\ {\cal D}^{\ast }_{A_{2}B_{2}}(u)\ = \ \d_{A_{1}A_{2}}\ 
\d_{B_{1}B_{2}}. \eqno (A.5)
$$
Again,
$$
\eqalignno
{
\br u_{1} \vert u_{2} \ke \ &= \ \sum_{A_{1}A_{2}}
\br A_{1} \vert A_{2} \ke \ {\cal D}^{\ast }_{A_{1}0}(u_{1})\ {\cal D}_{A_{2}0}(u_{2})\cr 
&= \ \sum_{A}{\cal D}^{\ast }_{A0}(u_{1})\ {\cal D}_{A0}(u_{2})\cr 
&= \ {\cal D}_{00}(u^{-1}_{1}u_{2}) .& (A.6) \cr 
}
$$ 
Thus,
$$
\eqalignno 
{
{\cal P}^{2}\ &= \ \int du_{1}du_{2}\ \vert u_{1} \ke \ \br u_{1}\vert 
u_{2} \ke \ \br u_{2} \vert \cr 
&= \ \int du_{1}du_{2}\ \vert u_{1} \ke \ {\cal D}_{00}(u^{-1}_{1}u_{2})\ 
\br u_{2} \vert .& (A.7) \cr 
}
$$
Let 
$$
u_{3} \ \equiv \ u^{-1}_{1}\ u_{2} . \eqno (A.8)
$$
Then, as $du_{1}\ = \ du_{3}$, (Haar Measure) and $\br u_{2} \vert 
\ = \ \br u_{1}u_{3}\vert $, we get
$$
\eqalignno 
{
{\cal P}^{2}\ &=\ \int du_{1}du_{3}\ \vert u_{1} \ke \ {\cal D}_{00}(u_{3})\ 
\br u_{1}u_{3} \vert \cr 
&=\ \int du_{1}du_{3}\ \vert u_{1} \ke \ {\cal D}_{00}(u_{3})\ 
\sum_{A} \br A \vert \ {\cal D}^{\ast }_{A0}(u_{1}u_{3}) \cr 
&=\ \int du_{1}du_{3}\ \vert u_{1} \ke \ {\cal D}_{00}(u_{3})\ 
\sum_{A_{1}A_{2}} \br A_{1} \vert \ {\cal D}^{\ast }_{A_{1}A_{2}}(u_{1})\ 
{\cal D}_{A_{2}0}(u_{3})\cr 
&=\ \int du_{1}\ \vert u_{1} \ke \ \sum_{A} \br A \vert \ {\cal D}^{\ast }_{A0}
(u_{1}) \cr 
&=\ {\cal P} . & (A.9) \cr 
}
$$  
\bigskip
\centerline{\bf Appendix B}
\bigskip
Let
$$
\hat{\r_{0}} \ \equiv \ \sum^{N-1}_{n=0} \vert n \ke \ \br n \vert .
\eqno (B.1)
$$
Further, we define
$$
e^{-\vert z \vert^{2}}\ \br \zbar \vert \ \hat{\r_{0}} \ \vert z \ke \ 
\equiv \ \r_{0}(\vert z \vert^{2}) . \eqno (B.2)
$$
Thus,
$$
\eqalignno 
{
\r_{0}(\vert z \vert^{2})\ &= \ e^{-\vert z \vert^{2}}\ \sum^{N-1}_{n=0}
\br \zbar \vert n \ke \ \br n \vert z \ke \cr 
&=\ e^{-\vert z \vert^{2}}\ \sum^{N-1}_{n=0}{{(\vert z \vert^{2})^{n}}
\over{n!}} . & (B.3) \cr 
}
$$
Now, let
$$
F(x,N) \ \equiv \ \sum^{N-1}_{n=0}{{x^{n}}\over{n!}}\ e^{-x} . \eqno (B.4)
$$
Then,
$$
\del_{x}F(x,N) \ = \ -e^{-x}\ {{x^{N-1}}\over{(N-1)!}}\ \simeq \ 
-e^{-x}\ {{x^{N}}\over{N!}} \eqno (B.5) 
$$
for large $N$.
Using Striling's formula
$$
N!\ \approx \ \sqrt{2\p }\ e^{-N}\ (N)^{N+{1\over{2}}} \eqno (B.6)
$$
and the saddle-point method
$$
e^{-x}\ x^{N} \ = \ e^{-(x-N\ln x)} \ \simeq \ e^{-(N-N\ln N)}\ 
e^{-{{(x-N)^{2}}\over{2N}}} \eqno (B.7)
$$
we get
$$
-\del_{x}F(x,N) \ \simeq \ {1\over{\sqrt{2\p N}}}\ 
e^{-{{(x-N)^{2}}\over{2N}}} . \eqno (B.8) 
$$
Therefore,
$$
\eqalignno 
{
-\del_{\vert z \vert^{2}}\r_{0}(\vert z \vert^{2})\ &\simeq \ 
{1\over{\sqrt{2\p N}}}\ e^{-{{(\vert z \vert^{2}-N)^{2}}\over{2N}}}\cr 
&=\ {1\over{\sqrt{2\p N}}}\ e^{-{{N(\vert \s \vert^{2}-1)}\over{2}}}\cr 
&\simeq \ {1\over{N}}\ \d ({{\vert z \vert^{2}}\over{N}}-1)\cr 
&=\ \d (\vert z \vert^{2}-N)  & (B.9) \cr 
}
$$
where $\s \equiv {{\vert z \vert^{2}}\over{N}}$.
Thus,
$$
\r_{0}(\vert z \vert^{2}) \ \approx \ \q (N-\vert z \vert^{2}).
\eqno (B.10)
$$
\bigskip
\centerline{\bf Appendix C}
\bigskip
>From (3.15), the exchange part of the Coulomb interaction is
$$
\eqalignno
{
C&={1\over{2}}\int d\vec k V(\vert \vec k \vert )\ \int d^2z_{1}\ 
d^2z_{2} e^{-\vert z_{1} \vert^{2} -\vert z_{2} \vert^{2}}
\vert \br \bar{z_{1}} \vert \rhoh \vert z_{2} \ke \vert^{2} \times \cr 
&\times  {\rm tr} \biggl[ P_{+} \bigl( u^{\dag } \ast e^{-i \chi(1)} \ast u \bigr) 
(z_{1}, \bar{z_{1}}) \ P_{+} 
\bigl( u^{\dag } \ast e^{i \chi(2)} \ast u \bigr)(z_{2}, \bar{z_{2}})
\biggr]. 
& (C.1) \cr 
}
$$

But,
$$
\eqalignno 
{
&\bigl( u^{\dag } \ast e^{-i \chi(1)} \ast u
\bigr)_{\b \g }(z_{1}, \bar{z_{1}})  = \delta_{\b \g } e^{-i \chi(1)}
- (\del_{\bar{z_{1}}}u^{\dag }u)_{\b \g }(\del_{z_{1}}e^{-i \chi(1)})
-(u^{\dag }\del_{z_{1}}u)_{\b \g }(\del_{\bar{z_{1}}}e^{-i \chi(1)}) \cr 
& +{1\over{2}}(\del^{2}_{\bar{z_{1}}}u^{\dag }u)_{\b \g }(\del^{2}_{z_{1}}
e^{-i \chi(1)}) + (\del_{\bar{z_{1}}}u^{\dag }\del_{z_{1}}u)_{\b \g }
(\del_{\bar{z_{1}}}\del_{z{1}}e^{-i \chi(1)}) 
+{1\over{2}}(u^{\dag }\del^{2}_{z_{1}}u)(\del^{2}_{\bar{z_{1}}}
e^{-i \chi(1)}) + \cdots .& (C.2) \cr 
}
$$

Similarly for 
$\bigl( u^{\dag } \ast e^{i \chi(2)} \ast u \bigr)(z_{2}, \bar{z_{2}})$.

Thus,
$$
\eqalignno
{
C&={1\over{2}}\int d^2z_{1}\
d^2z_{2} e^{-\vert z_{1} \vert^{2} -\vert z_{2} \vert^{2}}
\vert \br \bar{z_{1}} \vert \rhoh \vert z_{2} \ke \vert^{2}
\biggl[ {\rm tr} P_{+}V -{\rm tr} P_{+}(u^{\dag }\del_{z_{1}}u)
\del_{\bar{z_{1}}}V - h.c. \cr &-{\rm tr} P_{+}(u^{\dag }\del_{z_{2}}u)
\del_{\bar{z_{2}}}V - h.c. 
 +{\rm tr} P_{+}(u^{\dag }\del_{z_{1}}u)P_{+}
(u^{\dag }\del_{z_{2}}u) + h.c. \cr &+ (u^{\dag }\del_{z_{1}}u)P_{+}
(\del_{\bar{z_{2}}}u^{\dag }u)\del_{\bar{z_{1}}}\del_{z_{2}}V + h.c. 
+{\rm tr} P_{+}(\del_{\bar{z_{1}}}u^{\dag }\del_{z_{1}}u)\del_{\bar{z_{1}}}
\del_{z_{1}}V + 1 \rightarrow 2 + \cdots \biggr] . & (C.3) \cr 
}
$$ 

Let 
$$ z_{1} \equiv \sqrt{N} Z +{1\over{2}} z ; 
z_{2} \equiv \sqrt{N} Z -{1\over{2}} z \eqno (C.4)
$$
and
$$
u(\sqrt{N} Z, \sqrt{N}\bar{Z}) \equiv g(Z,\bar{Z}) .\eqno (C.5)
$$
Then,
$$
u(z_{1},\bar{z_{1}}) = g(Z,\bar{Z}) + {1\over{2\sqrt{N}}}z\del_{Z}
g(Z,\bar{Z}) + {1\over{2\sqrt{N}}}\bar{z}\del_{\bar{Z}}g(Z,\bar{Z})
+{1\over{4N}}\vert z \vert^{2}\del_{Z}\del_{\bar{Z}}g(Z,\bar{Z})
+\cdots .\eqno (C.6) 
$$
Again,
$$
\del_{z_{1}}u(z_{1},\bar{z_{1}}) = {1\over{\sqrt{N}}}
\del_{Z}\bigl[ g(Z,\bar{Z})
+{1\over{2\sqrt{N}}}z\del_{Z}g(Z,\bar{Z})
+{1\over{2\sqrt{N}}}\bar{z}\del_{\bar{Z}}g(Z,\bar{Z}) + \cdots 
\bigr] \eqno (C.7)
$$
$$
u(z_{2},\bar{z_{2}})=g(Z,\bar{Z}) - {1\over{2\sqrt{N}}}z\del_{Z}
g(Z,\bar{Z}) - {1\over{2\sqrt{N}}}\bar{z}\del_{\bar{Z}}g(Z,\bar{Z})
+{1\over{4N}}\vert z \vert^{2}\del_{Z}\del_{\bar{Z}}g(Z,\bar{Z})
+\cdots \eqno (C.8)
$$
$$
\del_{z_{2}}u(z_{2},\bar{z_{2}}) = {1\over{\sqrt{N}}}
\del_{Z}\bigl[ g(Z,\bar{Z})
-{1\over{2\sqrt{N}}}z\del_{Z}g(Z,\bar{Z})
-{1\over{2\sqrt{N}}}\bar{z}\del_{\bar{Z}}g(Z,\bar{Z}) + \cdots 
\bigr] \eqno (C.9)
$$
$$
\del^{2}_{z_{1}}u(z_{1},\bar{z_{1}})=\del^{2}_{z_{2}}u(z_{2},\bar{z_{2}})
={1\over{N}}\del^{2}_{Z}g(Z,\bar{Z}) \cdots  .\eqno (C.10)
$$

Further,
$$
\del_{z_{1}}V = -\del_{z_{2}}V = \del_{z}V \eqno (C.11)
$$
and thus,
$$
z\del_{z}V = \bar{z}\del_{\bar{z}}V = -{1\over{2}}V . \eqno (C.12)
$$

Again, from Appendix D, we get
$$
e^{-\vert z_{1} \vert^{2} -\vert z_{2} \vert^{2}}
\vert \br \bar{z_{1}} \vert \rhoh \vert z_{2} \ke \vert^{2}
=e^{-\vert z \vert^{2}}\bigl[ e^{{1\over{2\sqrt{N}}}(z\del_{z}-
\bar{z} \del_{\bar{z}})}\theta (1-\vert z \vert^{2})\bigr]
\bigl[ e^{-{1\over{2\sqrt{N}}}(z\del_{z}-
\bar{z} \del_{\bar{z}})}\theta (1-\vert z \vert^{2})\bigr]
. \eqno (D.13)
$$
Furthermore, $d^{2}z_{1}\ d^{2}z_{2} = N d^{2}Z\ d^{2}z $.

Therefore, using the above results, we obtain
$$
\eqalignno
{
C=&{{N}\over{2}}\int d^{2}z e^{\vert z \vert^{2}}\int d^{2}Z\ 
\bigl[ e^{{1\over{2\sqrt{N}}}(z\del_{z}-
\bar{z} \del_{\bar{z}})}\theta (1-\vert z \vert^{2})\bigr]
\bigl[ e^{-{1\over{2\sqrt{N}}}(z\del_{z}-
\bar{z} \del_{\bar{z}})}\theta (1-\vert z \vert^{2})\bigr]
\times \cr 
&\bigl[ V + {1\over{2N}}V{\rm tr}P_{+}g^{\dag }\del_{Z}\del_{\bar{Z}}g
+ h.c. + {1\over{N}}V{\rm tr}P_{+}\del_{\bar{Z}}g^{\dag }
\del_{Z}g \cr  
&- {2\over{N}}\del_{z}\del_{\bar{z}}V 
{\rm tr}P_{+}\bigl( g^{\dag }\del_{\bar{Z}}g g^{\dag }\del_{Z}g
-g^{\dag }\del_{\bar{Z}}g P_{+}g^{\dag }\del_{Z}g \bigr) \bigr] .
& (C.14) \cr 
}
$$  

The first term contains no excitations. Furthermore, the terms arising
from the derivatives of the theta functions are subleading in $N$.
Hence we drop these terms from further consideration.

Thus,
$$
\eqalignno
{
C& {1\over{8}}\simeq 
\int d^{2}z e^{-\vert z \vert^{2}}V \int d^{2}Z \theta (1-\vert 
Z \vert^{2})\bigl( 2 {\rm tr}P_{+}\del_{Z}(\del_{\bar{Z}}g^{\dag }g)
+ h.c. \bigr) \cr 
& -\int d^{2}z e^{-\vert z \vert^{2}}\del_{z}\del_{\bar{z}}V
\int d^{2}Z \theta (1-\vert
Z \vert^{2}){\rm tr}P_{+}\bigl( g^{\dag }\del_{\bar{Z}}g g^{\dag }\del_{Z}g 
-g^{\dag }\del_{\bar{Z}}g P_{+}g^{\dag }\del_{Z}g \bigr) . & (C.15) \cr 
}
$$

For $V={1\over{r}}$, 
$$
\int d^{2}z e^{-\vert z \vert^{2}}V = \sqrt{{{\pi }\over{2}}} {1\over{l}};
\int d^{2}z e^{-\vert z \vert^{2}}\del_{z}\del_{\bar{z}}V =
-{1\over{2}}\sqrt{{{\pi }\over{2}}} {1\over{l}} \eqno (C.16) 
$$
where $l\equiv {1\over{\sqrt{B}}}$ is the magnetic length. 

After performing the integral over $z,\bar{z}$, let us set
$Z \rightarrow {1\over{\sqrt{N}}} z$.

Then,
$$
\eqalignno
{
C = & - {1\over{2l}}\sqrt{{{\pi }\over{2}}}\int d^{2}z 
\theta (N - \vert z \vert^{2}) 
{\rm tr}P_{+} (g^{\dag }\del_{\bar{z}}g \ g^{\dag }\del_{z}g -
g^{\dag }\del_{\bar{z}}g P_{+} g^{\dag }\del_{z}g ) \cr 
&+{1\over{4l}}\sqrt{{{\pi }\over{2}}}\int d^{2}z 
\delta (N - \vert z \vert^{2}){\rm tr}P_{+} g^{\dag }
(\bar{z}\del_{\bar{z}}-z\del_{z})g . & (C.17) \cr 
}
$$
This is the expression quoted in (3.17) 
\bigskip
\centerline{\bf Appendix D}
\bigskip
We wish to prove:
$$
\br \bar{z_{2}} \vert \hat{f} \vert z_{1} \ke = e^{\bar{z_{2}} z_{1}}
\int d^{2}\omega d^{2}z e^{-\vert z \vert^{2}} e^{\bar{\omega }(z - z_{1})
-\omega (\bar{z} - \bar{z_{2}})} \br \bar{z} \vert \hat{f} \vert z \ke 
. \eqno (D.1) 
$$
The $R.H.S.$ can be reorganised as
$$
R.H.S. = \int d^{2}\omega e^{\bar{z_{2}} z_{1}}e^{-\bar{\omega }z_{1}
+ \omega \bar{z_{2}}}\int d^{2}z e^{-\vert z \vert^{2} + \bar{\omega }z
-\omega \bar{z}}f(z,\bar{z}) . \eqno (D.2) 
$$
But,
$$
\int d^{2}z e^{-\vert z \vert^{2} + \bar{\omega }z
-\omega \bar{z}}f(z,\bar{z}) = e^{-\vert \omega \vert^{2}}
e^{-\del_{\omega }\del_{\bar{\omega }}} f(-\omega , \bar{\omega }).
\eqno (D.3)
$$
Thus,
$$
R.H.S. = e^{\bar{z_{2}} z_{1}}\int d^{2}\omega e^{-\vert \omega \vert^{2}}
e^{-\bar{\omega }z_{1}
+ \omega \bar{z_{2}}} e^{-\del_{\omega }\del_{\bar{\omega }}}
f(-\omega , \bar{\omega }. \eqno (D.4)
$$
Let 
$$
\xi \equiv \omega + z_{1}; \bar{\xi } \equiv \bar{\omega } - \bar{z_{2}}.
\eqno (D.5)
$$
Then,
$$
\eqalignno
{
R.H.S. & = \int d^{2} \xi e^{-\vert \xi \vert^{2}} e^{-\del_{\xi }
\del_{\bar{\xi }}} f(-\xi +z_{1}, \bar{\xi }+ \bar{z_{2}}) \cr 
& = e^{\del_{z_{1}}\del_{\bar{z_{2}}}}\int d^{2} \xi 
e^{-\vert \xi \vert^{2}} e^{-\xi \del_{z_{1}}+\bar{\xi }\del_{\bar{z_{2}}}}
f(z_{1},\bar{z_{2}}) \cr 
&=e^{\del_{z_{1}}\del_{\bar{z_{2}}}}e^{-\del_{z_{1}}\del_{\bar{z_{2}}}}
f(z_{1},\bar{z_{2}}) \cr 
&=\br \bar{z_{2}} \vert \hat{f} \vert z_{1} \ke  \cr 
& = L.H.S . & (D.6) \cr 
}
$$

We further wish to show that:
$$
\int d^{2}\eta f(\eta , \bar{\eta }) \int d^{2}\omega 
e^{\bar{\omega }(\eta -z) -\omega (\bar{\eta }-\bar{z})}= f(z,\bar{z}) .
\eqno (D.7)
$$
Let 
$$
\chi \equiv \eta  -z . \eqno (D.8)
$$
Then,
$$
\bar{\omega }(\eta -z) -\omega (\bar{\eta }-\bar{z})= 2i(\omega_{x}
\chi_{y} - \omega_{y} \chi_{x}) . \eqno (D.9)
$$
Therefore, from $(D.7)$, 
$$
\eqalignno
{
L.H.S. &= \int d^{2}\eta f(\eta , \bar{\eta }) {1\over{\pi }}
\int d \vec \omega e^{2i(\omega_{x}
\chi_{y} - \omega_{y} \chi_{x})} \cr 
&=\int d\vec \eta f(\vec \eta ) \delta (\vec \eta -\vec z ) \cr 
&=f(z,\bar{z}) . & (D.10) \cr 
}
$$ 
\end